\documentclass{article}

\usepackage{arxiv}

\usepackage[utf8]{inputenc} 
\usepackage[T1]{fontenc}    
\usepackage{hyperref}       
\usepackage{url}            
\usepackage{booktabs}       
\usepackage{amsfonts}       
\usepackage{nicefrac}       
\usepackage{microtype}      
\usepackage{lipsum}
\usepackage{graphicx}
\usepackage{float}

\title{Investigation of infrasound noise background at Mátra Gravitational and Geophysical Laboratory (MGGL)}

\author{
  Edit Fenyvesi\\
  Institute of Particle and Nuclear Physics \\
  MTA Wigner Research Centre for Physics \\
  1121 Budapest, Konkoly Thege Miklós út 29-33.\\
  \texttt{fenyvesi.edit@wigner.mta.hu} \\
   \And
 József Molnár\\
 Laboratory of Electronics and Detector Development \\
  MTA Institute for Nuclear Research \\
  Hungary, 4026 Debrecen, Bem tér 18/c \\
  \texttt{jmolnar@atomki.mta.hu} \\
   \And
 Sándor Czellár\\
 Laboratory of Electronics and Detector Development \\
  MTA Institute for Nuclear Research \\
  Hungary, 4026 Debrecen, Bem tér 18/c \\
  \texttt{czellar.sandor@atomki.mta.hu} \\
}

\begin{document}
\maketitle

\begin{abstract}
Infrasonic and seismic waves are supposed to be the main contributors to the gravity-gradient noise (Newtonian noise) of the third generation subterranean gravitational-wave detectors. This noise will limit the sensitivity of the instrument at frequencies below 20 Hz. Investigation of its origin and the possible methods of mitigation have top priority during the designing period of the detectors. Therefore long-term site characterizing measurements are needed at several subterranean sites. However, at some sites, mining activities can occur. These activities can cause sudden changes (transients) in the measured signal, and increase the continuous background noise, too. We have developed a new algorithm based on discrete Haar transform to find these transients in the infrasound signal. We found that eliminating the transients decreases the variation of the noise spectra, and hence results a more accurate characterization of the background noise.  We also carried out experiments for controlling the continuous noise. Machines operating at the mine was turned on and off systematically in order to see their effect on the noise spectra. These experiments showed that the main contributor of the continuous noise is the ventilation system of the mine.
\end{abstract}


\section{Introduction}
In September 2015 the two aLIGO interferometers detected gravitational waves (GWs) for the first time in history \cite{Abbott(2016)}. The GW150914 observation was the first direct experimental proof of the existence of GWs that were predicted by Einstein on the basis of his general relativity theory. The GW was generated by a black hole-black hole merger. Since then the LSC-VIRGO collaboration published observations of GW signals from several more similar events (\cite{Abbott(2016.2)}\cite{Abbott(2017.2)}\cite{Abbott(2017.3)}\cite{Abbott(2017.4)}). The first observation of the GWs from merger of two neutron stars was on 17 August 2017 \cite{Abbott(2017)}. For the first time, the electromagnetic counterpart of the GW signal was detected by other laboratories, too \cite{Abbott(2017.5)}.

Gravitational-wave detectors help us to observe cosmic phenomena that can not be detected by other methods because their emitted electromagnetic radiation and the other cosmic ray particles could not reach the Earth. Detections of GWs and high-energy cosmic neutrinos have opened a new era of multimessenger astronomy  giving new insigths to the Universe.

aLIGO and AdVIRGO are expected to reach their design sensitivities in the next few years. However, even with their possible best performance, these detectors are not expected to be able to detect GWs generated by fast rearrangements in massive stellar bodies and in binary stellar systems with highly asymmetric mass distributions. Observation of these types of GW signals needs development of new concepts of GW detectors with increased sensitivity.

In the ideal case the relative distance of the test masses of a GW detector would only be changed by the GWs that pass through the interferometer. However, the relative movement of the test masses of the GW detectors is affected by the local environment, too. When extending the sensitivities to lower frequencies down to 1 Hz seismic and gravity-gradient noise are limiting. Gravity-gradient noise (Newtonian noise) is generated by changes of the density of matter near the detector. The density changes modify the local gravitational field around the detector and the resulting forces move the test masses. Therefore, the sensitivity of future GW detectors could be increased by installing them under the ground, where the changes of the density of the matter around the detector could be attenuated to a greater extent than on the Earth's surface.

A design study project of an underground GW detector has been proposed by eight European leading gravitational wave experimental research institutes \cite{Punturo(2011)}. Along with investigating the technical details of the instrument, site characterization experiments begun \cite {Beker(2015)} \cite{Barnafoldi(2017)}, too. The goals of the experiments are a) better understanding of the technical challenges that emerge during the process of building a detector under the ground and b) better understanding of the nature and origin of the underground environmental noises and, hence, the underground NN. The main aim of these experiments is selection of candidate sites for future GW detectors.

One of the site characterization experiments is performed at the Mátra Gravitational and Geophysical Laboratory (MGGL) \cite{Barnafoldi(2017)} that was established near Gyöngyösoroszi in Hungary in 2015. The laboratory is located in a cavern system of an unused ore mine 88 m below the surface. One of the aims of MGGL is studying the advantages of the underground installation of third generation gravitational wave detectors.

Specialized instruments have been installed in the laboratory for measuring the seismic and infrasound background and the electromagnetic noise. A long-time infrasound measurement program has been carried out to characterize the infrasound background noise in the laboratory.

Infrasound of the 0.01 - 10 Hz frequency range was monitored using a measuring system developed by HAS ATOMKI \cite{Van(2018)}. The results of the measurements were compared with the Bowman median noise model \cite{Bowman(2005)} that is generally used for characterization of the ambient atmospheric infrasound noise. For enabling the comparison we have developed a new signal processing pipeline, too.

During characterization of a potential site of a planned GW detector like ET, the investigation of the environmental background noises (seismic, electromagnetic, acoustic, etc.) is a fundamental step. However, there can be cases when mining or recultivation work is going on at mines near to potential candidate sites for future subterranean GW detectors. In order to separate transients from continuous background noise and separate noises of anthropogenic origin from noises of natural origin, new methods were developed \cite{Harms(2010)}.

In the mine where the MGGL laboratory was built, recultivation work is going on in three shifts per day. This activity causes sudden changes in the pressure of the air and transients in the measured infrasound signal, too. As we aimed characterization of the infrasound background noise with eliminating noise of local anthropogenic origin as much as possible, our data processing pipeline includes a step where the software finds the transients in the signal. Then data segments that contain the transients can be left out from further analysis.

In underground laboratories the air has to be ventilated and the ventilation system and other machinery can contribute to the infrasound background noise. In order to examine this, a controlled noise generating experiment was performed by us when turning off the ventilation system and a water pump for a short time was permitted by the operators of the mine.

The aims of this paper are a) presentation of the details of the software of the signal processing pipeline and b) results of the infrasound measurements obtained in MGGL in December 2018 and in the controlled noise experiment.

\section{Materials and Methods}

The infrasound detector of the measurement system is a new type of infrasound microphone (ISM1) developed by HAS Atomki. The data acquisition (DAQ) system is based on a Raspberry Pi3 Model B single-board computer that run Raspbian operating system. The analog signal of the infrasound microphone was connected to an Adafruit ADS1115 16-bit ADC. 

A real-time clock module was used to provide the time stamps with the appropriate accuracy. The measured data were stored on a 32 Gb microSD card connected to the Raspberry computer. The collected data were sent to a remote server that was operated outside of the mine in a building on the surface and provided internet access to the measured data. The downloaded data were processed off-line for further analysis.

The data processing pipeline is implemented in Python. In the followings the steps of the data processing are demonstrated using the data that were recorded in the 09 – 23 January 2019 period at MGGL with the infrasound monitoring system described above.

In the first step the downloaded data were divided into chunks of data measured in consecutive two-hour long time periods. Then the software searches each individual chunks for the possible transients.

Processing of each chunks began with the discrete Haar transform \cite{Bronstejn} of the signal $s_{i}^{(0)}$ , where $i=1,2,...,N$.
In each step of the transformation the approximation coefficients
$$s_{i}^{(n+1)}=\frac{1}{\sqrt(2)}\bigg(s_{2i-1}^{(n)}+s_{2i}^{(n)}\bigg)$$
and the detail components
$$d_{i}^{(n+1)}=\frac{1}{\sqrt(2)}\bigg(s_{2i-1}^{(n)}-s_{2i}^{(n)}\bigg)$$
were obtained. The approximation components show the trend, while the detail components show the fluctuations of the signal in different sub-bands of the signal.

In one or more sub-bands the detail components corresponding to transients in the original signal occur as outliers in the detail-coefficient vectors,
if the transients are relatively rare events.
Because of this, an outlier detection process had to be performed on each detail-coefficient vector individually.

First the measure of the spread of the components of a given detail-coefficient vector must be specified. The suitable measure can be influenced by outliers only in a small extent and, hence, it can be used as a reference that expresses the spread of the values without the outliers.

Therefore, the variance and the standard deviation are not suitable measures. They express how close are the observed data values to the mean value, however, the mean value can be influenced greatly by only a small number of outliers.
Considering this the median absolute deviation (MAD) was chosen as a measure of the spread \cite{MAD}.

Let's denote the median of the values of the k-th detail-coefficient vector with $m^{(k)}$.
Then the MAD of the n-th detail-coefficient vector is:
$$MAD^{(k)}=median(|d_{i}^{(k)}-m^{(k)}|)\quad.$$
If the distribution of the coefficients is a univariate normal one, the standard deviation would be $\sigma^{(k)}=1.4826 \cdot MAD^{(k)}$.
The distribution of $d_{i}^{(k)}$ values of a signal chunk without transients was expected to have a normal distribution approximately.
$d_{i}^{(k)}$ was considered as an outlier if $|d_{i}^{(k)}|> 5 \cdot 1.4826 \cdot MAD^{(k)}$.
If an outlier occurs at any level, the software finds the corresponding section at the original signal.

The data processing pipeline continued by dividing the two-hour long chunk into 128 sec long data segments.
Then the software checked whether a given segment included a transient. The check was done by searching for the sections found in the previous steps of the pipeline.
If the segment included (at least a part of) a transient, it was labeled as "noisy", else it was labeled as "silent".

The next steps are intended to get the so-called representative pressure amplitude spectral density (PASD) of the infrasound data collected.
PASD of data including "noisy" segments was compared to PASD of data that included only "silent" segments in order to see the effect of the transients of anthropogenic origin on the representative PASD.
Note that we aimed characterization of the infrasound background noise with eliminating noise of local anthropogenic origin as much as possible.

First the mean of each signal segment was subtracted from each value of the segment, then the signal was de-trended.
After that a Nuttall-window was applied on the signal: $\tilde s_{n}=w_{n} \cdot (s_{n}-\langle s \rangle)$.
Next, the windowed signal was transformed by fast Fourier transformation (FFT): $S_{k}$, where $k$ is the Fourier number related to the frequency $f=k \frac{f_{samp}}{N}$.
Then one-sided power spectral density(PSD) was computed:
$$PSD_{k}^{(s)}= \frac {2}{f_{samp} \cdot N \cdot W} \cdot |S_{k}| \quad,$$ 
where $W=\frac {1}{N} \sum_{n}^{N}w_{n}^{2}$ is a normalization factor that account for the power lost due to the windowing.
PSD expresses the power of the signal at a given frequency.

Pressure amplitude spectral density(PASD) is the square root of the PSD:
$$PASD_{k}^{(s)} = \sqrt {PSD_{k}^{(s)}}\quad.$$

By choosing PASD, the infrasound noise level can be expressed with the unit of pressure (Pa) per $\sqrt{Hz}$.
To characterize infrasound background noise of the site, for each $k$, the median and the 10th and 90th percentile of all segment's $PASD_k^{(s)}$ values were computed.
The 10th (90th) percentile shows the value below which $10\%$ ($90\%$) of the observations may be found, and hence is suitable to inform us about the variations in the PASD.
\section{Results and discussion}

\subsection{Effect of removing data segments with transients on the representative pressure amplitude spectral density(PASD)}
\unskip

In the case of MGGL the anthropogenic noises are generated by the different activities of the recultivation of the mine. Some systems of the mine’s infrastructure (e.g. elevators, ventilation system, water pumps, etc.) generate continuous background noise that is always present. Furthermore, other activities (e.g. opening and closing sluices) occasionally can cause sudden changes in the air pressure and hence transients in the measured infrasound signal.

Figure 1 shows results of the infrasound background measurements performed at MGGL in the 09 – 23 January 2019 period when there was recultivation activity at the site of the mine. The red dashed line, the red solid line and the red dashed-dotted line indicate the 90th percentile, the median and the 10th percentile of the measuerd PASD before the filtering procedure, respectively. The blue lines indicate the 90th percentile, the median and the 10th percentile of the PASD, that was obtained after carrying out the filtering procedure described above in the Materials and Methods section. It can be seen that the curves are very similar but the 90th percentile curves are significantly different below 0.2 Hz.

It has to be emphasized that the PASD curves obtained without the application of the filtering procedure do not inform us on the origin of the low frequency parts of the curves. In other words, further information is needed to decide if the low frequency parts of the curves are the results of the increase of the continuous low frequency noise background or transients contribute, too.
The difference of the 90th percentile curves is caused by the application of the filtering procedure. This means, that comparison of the PASDs obtained with and without the filtering procedure is a useful method for obtaining information on the frequency of the occurence of transients in the background noise.

\begin{figure}[H]
\centering
\includegraphics[width=14 cm]{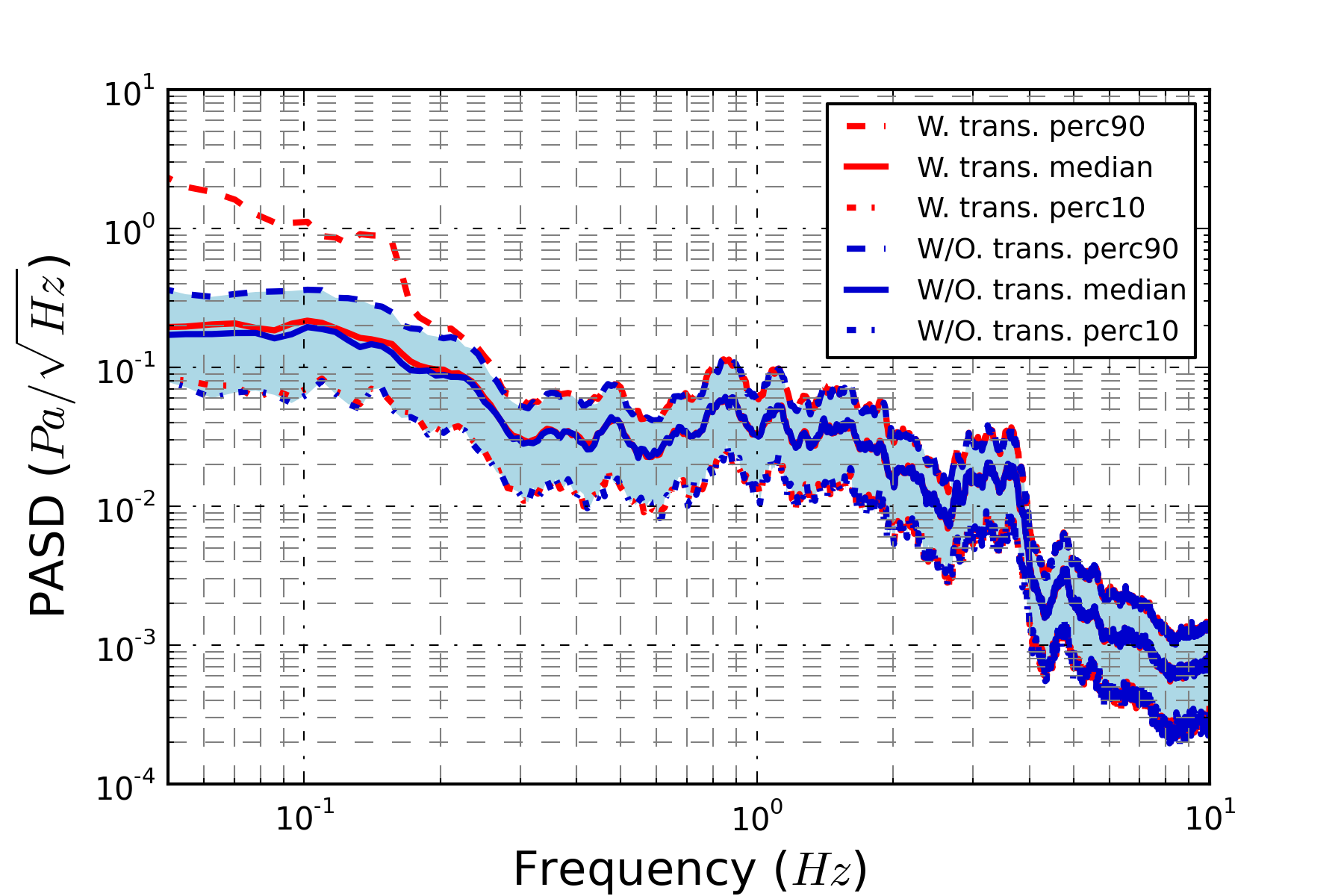}
\caption{The red lines show the PASD curves computed from measurement data containing transients, while the blue lines correspond to data without transients.}
\label{MGGL19jan}
\end{figure}

\vspace{6pt}
\subsection{Investigation of the sources of the continuous background infrasound noise at MGGL and comparision with the Bowman models}

On 28 December 2018 a controlled experiment was carried out, too, for studying the effects of noise sources on the infrasound noise background of MGGL. Miners suspended the recultivation work on that day. As the first step of the experiment a water pump near the laboratory was turned off. Then a door that separates the corridor of MGGL and a vertical shaft (that is opened to the surface) was opened. In the third step the ventilation system was turned off, too. During the experiment the infrasound noise background was monitored.
Figure 2 shows, for comparison, the median curve of the PASD (dark blue line) that was obtained from the data recorded in the measurement course in the 09-23 January 2019 period. The PASD curve is considered as the reference for the normal working days at the mine. Also, Figure 2 shows the mean curves of the PASDs obtained for the three steps of the experiment.
As the light blue curve in Figure 2 shows, the turning off of the water pump barely affected the infrasound noise background. After opening the door the noise level increased significantly in the f = 0.4 Hz – 0.7 Hz and the f = 1 Hz – 2 Hz frequency ranges (see the purple curve in Figure 2). The main reason for the changes was the strong air flow in the vertical shaft. The air flow vibrates the walls of the shaft, too, and the vibration waves can propagate to the walls of the MGGL contributing to the noise level in the laboratory even when the door is closed. The orange curve in Figure 2 clearly shows that the noise level decreased very significantly practically in the whole f = 0.05 Hz – 10 Hz frequency range after turning off the ventilation system. This means that the main source of the continuous noise that is observable in MGGL is the ventilation system.

\begin{figure}[H]
\centering
\includegraphics[width=14 cm]{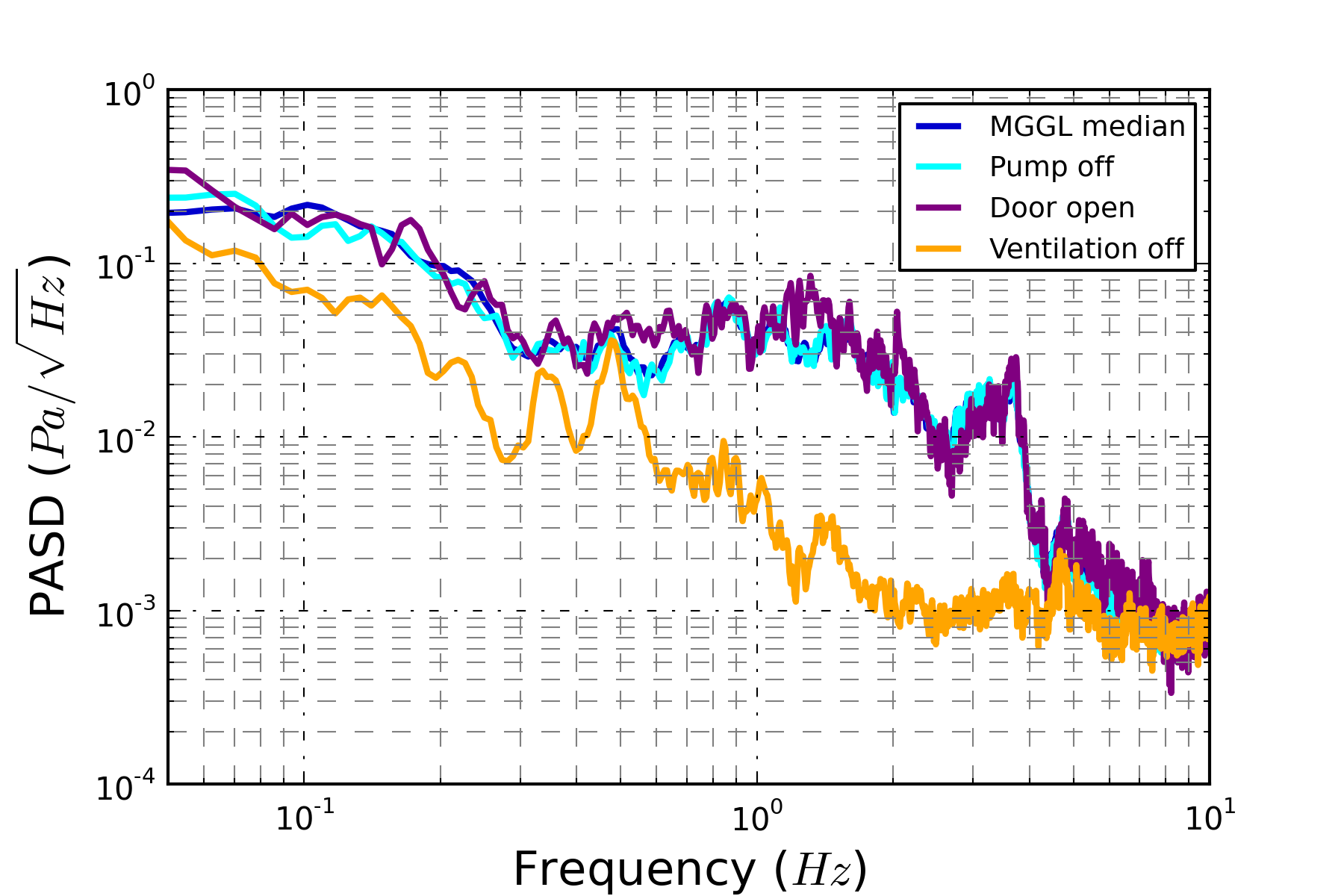}
\caption{Results of noise controlling experiment. The MGGL-median curve corresponds to the measurement campaig of January, 2019.}
\label{silentDay}
\end{figure}

\vspace{6pt}
Figure 3 shows the frequency dependency of the median of the PASD (orange solid line) that was obtained from our measurements performed in MGGL when the ventilation system was turned off. Also, Figure 3 shows the three curves (low noise model, median, high noise model) that were obtained from the models developed and published by Bowman et al. in \cite{Bowman(2005)} for characterization of the global ambient infrasound noise background on Earth’s surface. Bowman et al. \cite{Bowman(2005)} derived their three models from the data measured at many infrasound measurements stations around the world.
One can see that the median of the PASD at MGGL agrees well with the median of the Bowman model in the 0.05 Hz - 0.15 Hz region and there are differences above 0.15 Hz and the trend of our measurements is similar to the trend of the Bowman model median.
	Our results show the importance of the careful acoustic planning of the human activity and the infrastructure, especially the ventilation systems and the connections of the tunnels to the surface of the site, in the cases of future underground GW detectors.

\begin{figure}[H]
\centering
\includegraphics[width=14 cm]{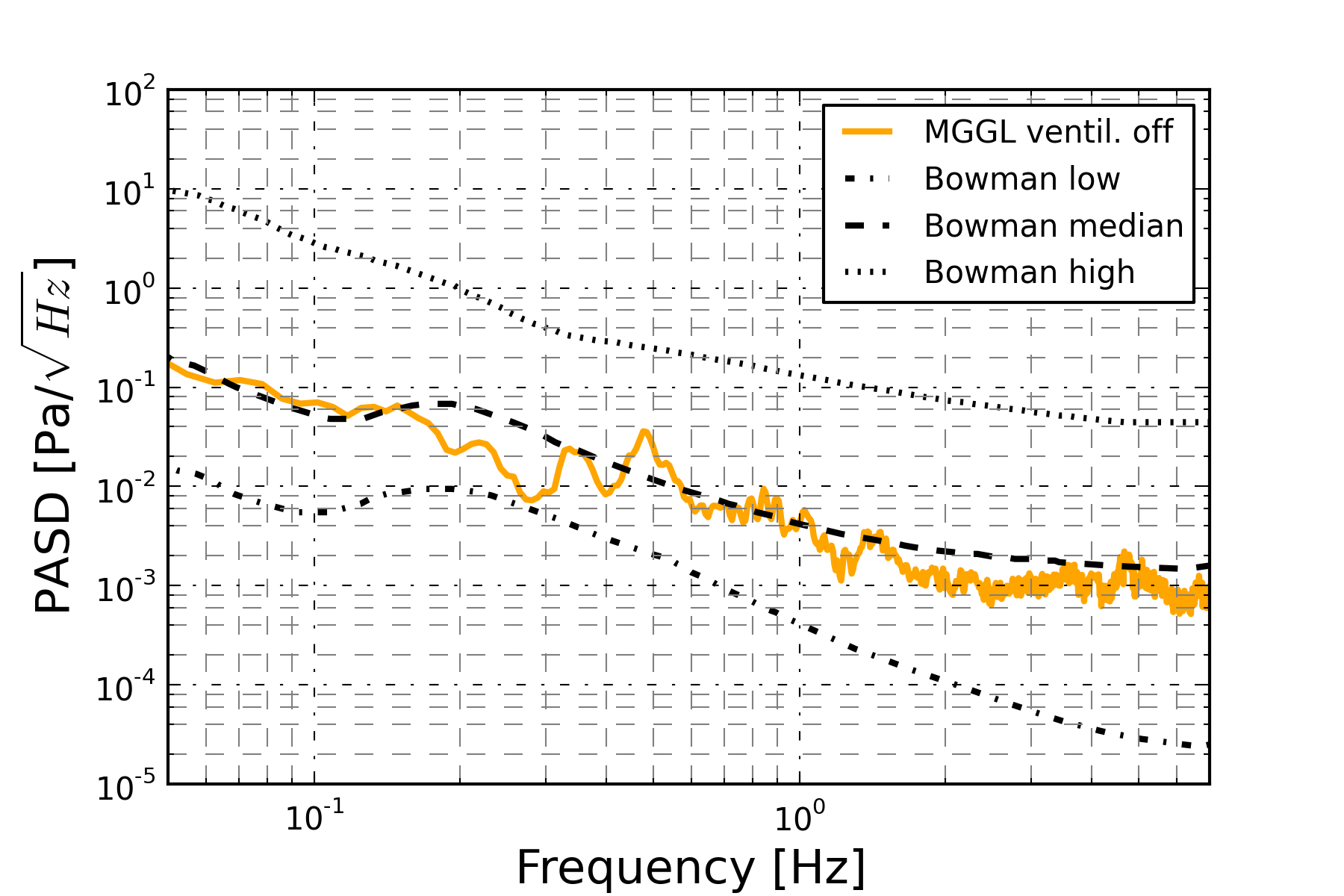}
\caption{The orange line indicates the lowest noise level measured at MGGL, when the ventilation system was turned off.}
\label{ventBow}
\end{figure}
\section{Conclusions}

A method and a data processing pipeline software have been developed for analyzing the infrasound intensity data series measured studying the infrasound noise background at the Mátra Gravitational and Geophysical Laboratory (MGGL). The discrete Haar transform is employed for identification of the infrasound noises from anthropogenic activities around the MGGL. An experiment has been carried out, too, when there was no anthropogenic activity in the MGGL and in the cavern system around it. The frequency spectra and other quantities obtained from the analysis of the measured data have been used for the site characterization of MGGL and for studying its suitability for hosting the internationally proposed Einstein Telescope gravitational wave detector.

\section*{Acknowledgement}

The support of the European Regional DevelopmentFund and Hungary in the frame of the project GINOP-2.2.1-15-2016-00012 is acknowledged. The work was supported by the grants National Research, Development and Innovation Office –NKFI 124366. The contribution and support of Nitrokemia Zrt. in particular Á. Váradi and V. Rofrits is acknowledged. We also thank the construction work of Geofaber Zrt. The authors thank Géza Huba, Péter Ván, Mátyás Vasúth and László Somlai for the constant support, too.

\end{document}